\documentclass{article}

\usepackage{graphicx}
\usepackage{etex}

\usepackage{listings}
\usepackage{standalone}

\usepackage{bmpsize}

\usepackage[table,xcdraw]{xcolor}
\usepackage{tikz}

\usepackage{booktabs}
\usepackage{multirow}

\usepackage{float}
\restylefloat{table}
\usepackage{amssymb}
\usepackage{tikz}

\usepackage{pbox}
\usepackage{amssymb}
\usepackage[document]{ragged2e}

\usepackage{amsmath,amsfonts,amssymb,amsthm,epsfig,epstopdf,url,array}

\usepackage{makeidx}
\usepackage{nomencl}

\usetikzlibrary{shapes.geometric, arrows}

\usepackage[T1]{fontenc}
\usepackage[latin9]{inputenc}
\usepackage{amsmath}
\usepackage{xargs}[2008/03/08]

\usepackage[english]{babel}

\usepackage{longtable}

\usepackage{afterpage}

\usepackage{pdfpages}
\usepackage{lipsum}% just to generate filler text
\usepackage{hyperref}
\usepackage{caption}
\usepackage{latexsym}
\usepackage{amssymb}

\usepackage{mathrsfs}

\usepackage{xspace}
\usepackage{graphics}
\usepackage{tikz}
\usepackage{mathtools}
\usepackage{bussproofs}
\EnableBpAbbreviations
\newcommand{\UICm}[1]{\UIC{$#1$}}
\newcommand{\AXCm}[1]{\AXC{$#1$}}
\newcommand{\BICm}[1]{\BIC{$#1$}}

\newcommand{\RLm}[1]{\RL{$#1$}}

\def\land{\wedge}

\usepackage{setspace}

\usepackage[T1]{fontenc}
\usepackage{textcomp}
\usepackage{listings}

\usepackage{amssymb}

\newcommand{\Orefl}{
    \AXCm{~} \RLm{\text{refl}\, t}
    \UICm{\vdash t = t} 
    \DP}

\newcommand{\Oassume}{
    \AXCm{~} \RLm{\text{assume} \, \varphi}
    \UICm{\left\{{\varphi}\right\} \vdash \varphi} 
    \DP}

\newcommand{\Oeqmp}{
    \AXCm{\Delta \vdash \varphi}
    \AXCm{\Gamma \vdash \varphi = \psi}
    \RLm{\text{eqMp}}
    \BICm{\Gamma \cup \Delta \vdash \psi} 
    \DP}

\newcommand{\Oabs}{
    \AXCm{\Gamma \vdash t = u} \RLm{\text{absThm} \, v}
    \UICm{\Gamma \vdash (\lambda v.t) = (\lambda v. u)}
    \DP}

\newcommand{\Oapp}{
    \AXCm{\Gamma \vdash f = g}
    \AXCm{\Delta \vdash x = y}
    \RLm{\text{appThm}}
    \BICm{\Gamma \cup \Delta \vdash f x = g y}
    \DP
}

\newcommand{\Oda}{
 \AXCm{\Gamma \vdash \varphi}
    \AXCm{\Delta \vdash \psi}
    \RLm{\text{deductAntisym}}
    \BICm{(\Gamma \setminus \left\{{\psi}\right\} )\cup (\Delta \setminus \left\{{\varphi}\right\}) \vdash \varphi = \psi}
    \DP
}

\newcommand{\Osubst}{
    \AXCm{\Gamma \vdash \varphi} \RLm{\text{subst} \sigma}
    \UICm{\Gamma [\sigma] \vdash \varphi[\sigma]} 
    \DP}

\newcommand{\Obetaconv}{
    \AXCm{~} \RLm{\text{betaConv} ((\lambda v.t) u)}
    \UICm{\vdash (\lambda v.t) u = t[u/v]} 
    \DP}
\newcommand{\OdefineTypeOp}{
    \AXCm{\vdash \varphi \, t}
    \RLm{\text{defineTypeOp} \enspace n\enspace abs\enspace rep\enspace vs}
    \UICm{\vdash abs (rep \, a) = a\enspace\enspace\enspace\enspace \vdash  (rep (abs\, r) = r) = \varphi \, r}
    \DP
}

\newcommand{\OdefineConst}{
    \AXCm{~}
    \RLm{\text{defineConst} c t}
    \UICm{c = t}
    \DP
}

\newcommand{\OproveHyp}{
    \AXCm{\Gamma \vdash \varphi}
    \AXCm{\Delta \vdash \psi}
    \RLm{\text{proveHyp}}
    \BICm{\Gamma \cup (\Delta \setminus \left\{ {\varphi}\right\}) \vdash \psi}
    \DP
}

\newcommand{\Otrans}{
 \AXCm{\Gamma \vdash s = t} \AXCm{\Delta \vdash t = u} \RLm{\text{trans}}
    \BICm{\Gamma \cup \Delta \vdash s = u}
\DP
}

\newcommand{\Osym}{
\AXCm{\Gamma \vdash \varphi = \psi}
\RLm{\text{sym}}
\UICm{\Gamma \vdash \psi = \varphi}
\DP
}

\newcommand{\OdefineTypeOpnew}{
    \AXCm{\vdash \varphi \, t}
    \RL{defineTypeOp\, n\, abs\, rep\, vs}
    \UICm{\vdash abs (rep \, a) = a \enspace\enspace\enspace\enspace \vdash \lambda r. \varphi \, r = (\lambda r. (rep (abs\, r) = r)) }
    \DP
}

\newcommand{\vbar}{$\biggr\rvert$}

\newcommandx\prove[3][usedefault, addprefix=\global, 1=]{\AXCm{#2}\RL{#1}\UICm{#3}\DP}

\newcommandx\translate[1][usedefault, addprefix=\global]
{{\vbar}{#1}{\vbar}}

\usepackage{multicol}

\usepackage{lscape}

\newcommand{\examplehol}{
\begin{prooftree}

\AXCm{}
\RLm{ASSUME}
\UICm{p \land  (p\Rightarrow q) \vdash p \land (p\Rightarrow q)}
\RLm{CONJUNCT2}
\UICm{p \land  (p\Rightarrow q) \vdash p\Rightarrow q}

\AXCm{}
\RLm{ASSUME}
\UICm{p \land  (p\Rightarrow q) \vdash p \land (p\Rightarrow q)}
\RLm{CONJUNCT1}
\UICm{p \land  (p\Rightarrow q) \vdash p}

\RLm{MP}
\BICm{p \land  (p\Rightarrow q) \vdash q} 
\RLm{DISCH}
\UICm{\vdash (p \land (p \Rightarrow q)) \Rightarrow q}

\DP
}

\usepackage{arxiv}
\usepackage{tabularx} 
\usepackage[]{inputenc} % 
\usepackage{multirow}

\title{ProofCloud: A Proof Retrieval Engine for Verified Proofs in Higher Order Logic}

\author{
 Shuai Wang \\
 INRIA Rocquencourt, Paris, France\\
ILLC, University of Amsterdam, The Netherlands\\
  \texttt{shuai.wang.vu@gmail.com} \\
}

\begin{document}
\maketitle
\begin{abstract}
This paper introduces ProofCloud, a proof retrieval engine for verified proofs in higher order logic. It provides a fast proof searching service for mathematicians and computer scientists for the reuse of proofs and proof packages. In addition, it includes the first complete proof-checking results and benchmarks of the OpenTheory repository.\footnote{The paper has been presented at the International Workshop on User Interfaces for Theorem Provers (UITP) in 2016.}
\end{abstract}

% keywords can be removed

\section{Introduction}
\justify

Interactive Theorem Provers (ITPs)  have been playing an important role in formal mathematics, software verification and hardware verification. In recent years, there has been dramatic progress in the usability and power of ITPs. Among them, theorem provers in the HOL family are among the most widely used in this domain. The HOL family consists of HOL Light \cite{hollight}, HOL4 \cite{hol4} and ProofPower \cite{proofpower}, to name a few. These ITPs implement the same higher order logic, namely Church's simple type theory \cite{opentheory}. However, they each contain significant theory formalisations that are not accessible to each other. HOL Light \cite{hollight} has a formalisation of complex analysis while HOL4 has a formalisation of probability theory \cite{hol4}. In contrast, ProofPower has a formalization of the Z specification language \cite{proofpower}. These ITPs were developed without considering proof sharing and reuse to avoid repeated work proving large theorems. Moreover, ITPs may not be bug-free and may lead to errors in generated proofs which are not necessarily apparent within the proof systems themselves. Additionally, proofs can be huge, making them difficult or even impossible to be checked by hand (e.g. the Kepler Conjecture \cite{kepler}). The demand of reliability of such systems leads to the necessity of proof checking, especially methods and tools independent from the ITPs involved. All this requires theorem provers to export proofs in a certain format. OpenTheory is a format and a proof repository for the HOL family with this purpose. Taking advantage of the similarity of the logic and design between these systems, OpenTheory \cite{opentheory} has developed a standard format for serializing proofs and sharing proofs \cite{opentheory}. ITPs in the HOL family can export proofs to the OpenTheory format as article files (packages of proofs and relevant information). These files are then converted to Dedukti \cite{dedukti} format by a proof translator, namely Holide \cite{holide}, for the sake of proof checking.

% Theorem provers in the HOL family are based on higher order logic. 
HOL Light \cite{hollight} is an open source interactive theorem prover for higher order logic.  Its logic is an extension of Church's simple type theory with polymorphic type \cite{hollight}.  Higher order logic is also known as simple type theory. It is a logic on top of simply typed $\lambda$-calculus with additional axioms and inference rules \cite{hollight}. The type of a term is either an individual, a boolean type or a function type. A term is either a constant, a variable (e.g. $x$), an abstraction (e.g. $\lambda x.x$) or a well-typed application (e.g. $(\lambda x.x) y)$. The notation $x : \iota$ means that the term $x$ is of type $\iota$. Types are sometimes omitted for simplicity of representation. The equality is of polymorphic type and plays three roles in HOL Light: definition, equivalence and bi-implication.

\begin{table}[]
\centering
\begin{tabular}{ll}
type variables & $\alpha, \beta$ \\ 
type operators  & $p$   \\
types & $A, B::= \alpha \,|\, p(A_{1},\ldots,A_{n})$ \\
term variables & $x, y$\\
term constants  & $c$ \\
terms & $M, N ::= x \,|\, \lambda x: A. M \,|\, M N \,|\, c$ \\
\end{tabular}
\end{table}

% The kernel of HOL Light is an OCaml file where terms, types, symbols and inference rules are defined. Symbols and inference rules in the kernel are considered primitive. On top of the kernel, additional symbols are introduced and inference rules are derived. 

This paper presents ProofCloud, a proof retrieval engine for retrieval of verified higher order logic proofs and some related proof checking results. This paper is organised as follows: we first introduce OpenTheory and explain its kernel and the recent updates in Section \ref{opentheory}. Following that is ProofCloud (Section \ref{ch:proofcloud}) and the implementation (Section \ref{impli}). Finally, the evaluation and benchmarks of proof checking are included.

% \section{Higher Order Logic}
% \label{ch:kernel}

\section{OpenTheory}
\label{opentheory}

% \subsection{Higher Order Logic}

% \subsection{OpenTheory}

OpenTheory \cite{opentheory} is a cross-platform proof package manager for proofs in ITPs of the HOL family. The standard library of OpenTheory groups theorems into packages, including theorems of booleans, sets, lists, etc. Moreover, it contains a repository of proof packages allowing sharing and reuse of proofs between theorem provers in the HOL family. OpenTheory has inspired further exploration of proof management between HOL families \cite{kaliszyk2013scalable} as well as the development of several projects (e.g. \cite{holide} and \cite{assaf2015mixing}) taking advantage of these packages. 
\begin{table*}[!ht]
\centering
\renewcommand{\arraystretch}{2.5}
\caption{OpenTheory version 5 Logic Kernel }
\label{tb:opentheoryrules}
\begin{tabular}{l|llll|l}

\multicolumn{2}{l}{\Orefl} & \multicolumn{2}{l}{\Oassume} & \multicolumn{2}{l}{\Oeqmp} \\
\multicolumn{3}{l}{\Oabs}             & \multicolumn{3}{l}{\Oapp}             \\
\multicolumn{4}{l}{\Oda}                         & \multicolumn{2}{l}{\Osubst} \\
\multicolumn{4}{l}{\Obetaconv}                         & \multicolumn{2}{l}{\OdefineConst} \\
\multicolumn{6}{l}{\OdefineTypeOp}                                                 \\ 
\end{tabular}
\end{table*}

Table \ref{tb:opentheoryrules} illustrates the primitive inference rules of the OpenTheory logic kernel which has some small differences compared with the HOL Light kernel\footnote{Note that $appThm$ Corresponds to MK\_COMB in HOL Light.}. Different from other proof libraries, OpenTheory was not designed to be a general purpose proof repository. It is a specialised library and format for proof sharing between the ITPs in the HOL family.

\section{ProofCloud}
\label{ch:proofcloud}
% Todo. update the size of "1K proofs".

 We introduce ProofCloud\footnote{\url{http://airobert.github.io/proofcloud/}}, a proof retrieval engine, for easy searching of verified higher order logic proofs. It consists of a presentation of proofs and proof packages, the results of some statistical and structural analyses, together with their proof checking results. So far it has been populated with 1687 proofs from 6 packages of OpenTheory.  Information of each proof and package is presented on seperate webpages. Taking classical proofs as those using the axiom of choice, ProofCloud can distinguish between classical proofs and constructive proofs. Furthermore, it tracks the origin of classicism, in other words, which classical lemmas\footnote{To avoid confusion, all theorems used to prove the conclusion are referred to as lemmas when referring to a specific theorem.} were used within the proof. With this ability, the amount as well as the percentage of classical proofs of a certain package is calculated and displayed on its (package) page. Users will also find a statistical analysis on the size of proofs and links to proof checking results.  As far as the author knows, this is the only online proof retrieval engine of its kind. The specifications of ProofCloud is presented in Appendix \ref{spec}.
 
 \begin{figure}[]
\caption{The Interface of ProofCloud}
\centering
\includegraphics[width=0.95\textwidth]{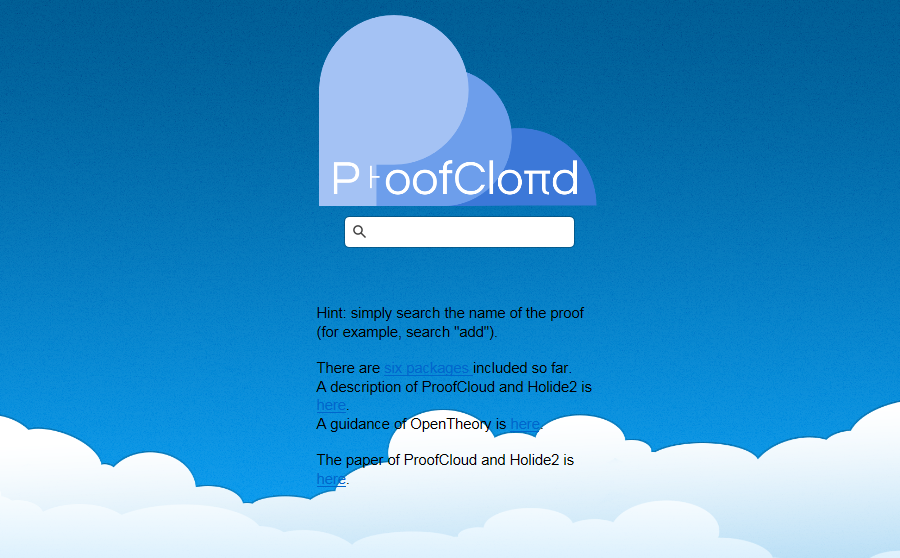}
\end{figure}

\section{Implementation}
\label{impli}
\begin{figure}[]
\caption{A Proof Page of ProofCloud}
\centering
\includegraphics[width=0.95\textwidth]{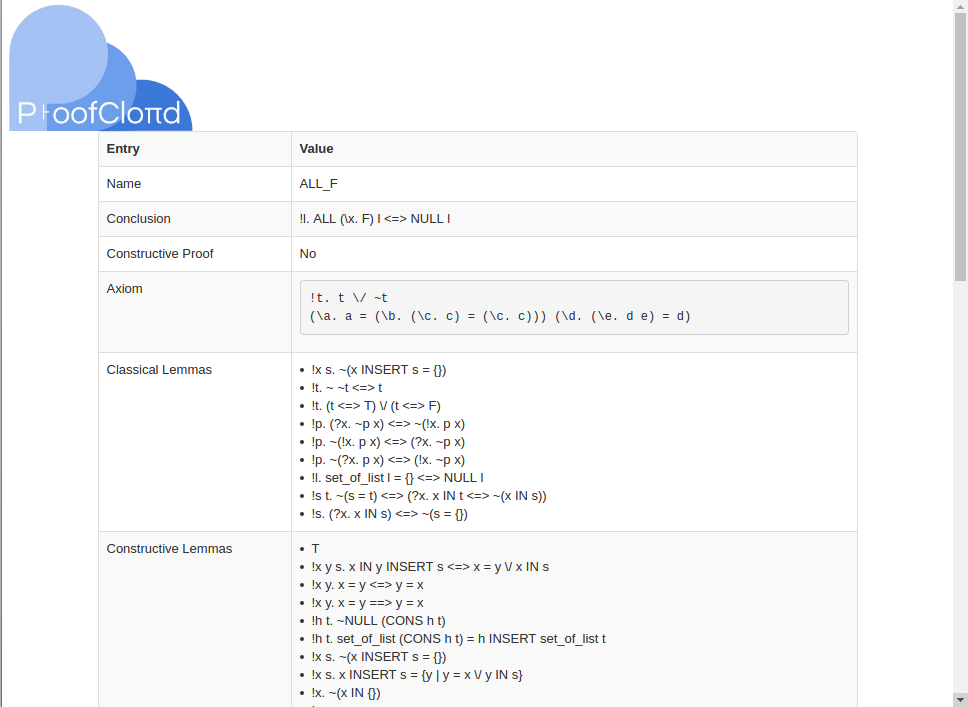}
\label{aproof}
\end{figure}

ProofCloud is a customised search engine from Swiftype \footnote{Swiftype (\url{https://swiftype.com/}) is a customisable search engine with simple APIs and real-time indexing.}. It was implemented to provide fast searching service of proofs for mathematicians and computer scientists using ITPs in the HOL family.  It has an elegant and user-friendly interface with the search results clearly displayed. It has been on service since July 2015 and just been upgraded to version 2. It contains more than 1700 pages of information about theorems and packages of higher order logic. These pages were generated by extending a modified version of HOL Light\footnote{\url{http://src.gilith.com/hol-light.html}} with functions to import and export proof articles and generate proof analysis results in the form of webpages (for example Figure \ref{aproof}). More specifically, proof analysis takes advantage of the proof logging method. Instead of erasing the proof after exporting, it performs proof analysis and generates webpages with a representation of the results.

% ProofCloud takes advantage of packed proofs in OpenTheory repository and record proof checking results by Holide and Dedukti.

% \section{The Dependency of Packages of OpenTheory}
% \label{dependency}

\begin{figure}[!ht]
 \centering
\begin{tikzpicture}[scale=0.8]

% \caption{Proof Dependency Analysis}
\tikzstyle{conjecture}=[rectangle]
\tikzstyle{title} = [rectangle]
% \small
\node[conjecture] (base) at (0, -1) {base (the standard library)};
\node[conjecture] (stream) at (-2, 0.5) {stream};
\node[conjecture] (divides) at (2, 0.5) {natural-divides};
\node[conjecture] (prime) at (0, 2) {natural-prime};
\node[conjecture] (fibonacci) at (-4, 3) {natural-fibonacci};
\node[conjecture] (modular) at (4, 3) {modular};
\node[conjecture] (probability) at (-1, 4) {probability};
\node[conjecture] (gfp) at (-5, 5) {gfp};
\node[conjecture] (list) at (-1, 6) {natural-list};
% \node[conjecture] (word) at (4, 6) {word}
% \node[conjecture] (word10) at (0,7) {word10}

\draw[->] (stream) -- (base); 
\draw[->] (divides) -- (base);
\draw[->] (prime) -- (divides);
\draw[->] (prime) -- (stream);
\draw[->] (fibonacci) -- (stream);
\draw[->] (probability) -- (stream);
\draw[->] (gfp) -- (fibonacci);
\draw[->] (list) -- (probability);
\draw[->] (modular) -- (divides);
% \draw[->] (word) -- (modular);
% \draw[->] (word10) -- (word);
% \draw[->] (word10) -- (list);
% \draw[->] (bot1) -- (forall1);

\end{tikzpicture}
\caption{Dependency of Packages of OpenTheory}
\label{fg:logicdependency}
\end{figure}
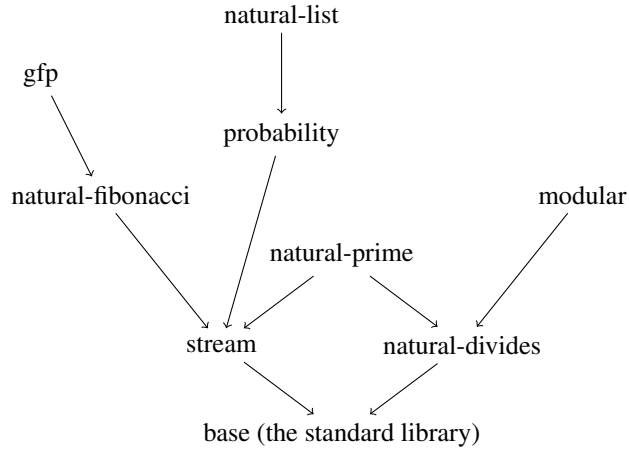

So far ProofCloud includes six OpenTheory packages in total. These packages are \textit{base} (the standard library with some subpackages), \textit{stream}, \textit{probability}, \textit{natural-bits}, \textit{natural-divides} and \textit{natural-prime}. Other packages will be added in the near future. An analysis of the dependency of packages of OpenTheory is as shown in Figure \ref{fg:logicdependency}. To load a package, it is required to load all the packages it depends on primarily. In addition, the package \textit{modular} is a parametric theory and requires the \textit{modular-witness} theory which defines a suitable signature. Similarly \textit{gfp} is another parametric theory. These packages are being updated and therefore not included for now. The package \textit{modular} and \textit{gfp} and packages depending on them will be added to ProofCloud in the near future.

% \begin{center}
% \end{center}

\subsection{Updates of OpenTheory}

To obtain the proof checking results for ProofCloud, Holide has to be updated according to the changes of OpenTheory. In version 6\footnote{A full comparison table between version 5 and version 6 is online at \url{http://www.gilith.com/pipermail/opentheory-users/2014-December/000461.html}}, OpenTheory expanded its logic kernel with four inference rules: \textit{proveHyp}, \textit{trans}, \textit{sym} and \textit{defineTypeOp} (as in Table \ref{opentheory6rules}) to generate more compact article files (packages of proofs and some relevant information). Following this update, some projects depending on the OpenTheory repository are out of date. Holide (version 1) \cite{holide} is one of them \footnote{The old Holide (version 1) is at \url{https://www.rocq.inria.fr/deducteam/Holide/}. The upgraded version is maintained by the author and is available at \url{http://airobert.github.io/holide/} also as an open-source software.}.

\begin{table*}[!ht]
\centering
\caption{New and Updated Inference Rules of OpenTheory (version 6)}
\label{opentheory6rules}
\begin{tabular}{l}
\OproveHyp\\ 
\Otrans \\
\Osym \\
\OdefineTypeOpnew 
\end{tabular}
\end{table*}

\subsection{Holide, Dedukti and Proof Checking}
\label{ch:holide}

  Dedukti is a logical framework for defining logics and expressing proofs. Cousineau and Dowek proved that Higher Order Logic can be embedded in the $\lambda\Pi$-calculus Modulo \cite{gilles1}. This resulted in proposing Dedukti as a proof checker \cite{dedukti}. To transform proofs from the OpenTheory format \cite{opentheory} to the Dedukti format, we need a translator, namely Holide \cite{holide}. Holide uses a modular translation of higher order logic to Dedukti, making it possible to extend the translation \cite{holide}. Recent updates of the OpenTheory urges an update of the Holide program. The following shows an extension of the bijective translation by adding corresponding translation of the \textit{sym}, \textit{trans} and \textit{proveHyp} inference rules. Moreover, \textit{defineConstList} and \textit{hdTl} are added while processing the proof articles. In addition, the translation of \textit{defineTypeOp} is updated while the commands \textit{version} and \textit{pragma} are omitted during the translation. The translation notations of types, terms and proofs are kept consistent with \cite{holide}. On the left are the (OpenTheory) inferences and on the right are the corresponding Dedukti formula. Details of translation are as follows:

\begin{itemize}
\item[] \translate{\Osym} = ${Sym|A|}|\varphi||\psi||D_{1}|$, where $\varphi$ and $\psi$ are of type $A$ and $D_1$ is the proof of $\varphi = \psi$.
\item[] \translate{\Otrans} = \\${Trans|A||s||t||u||D_{1}||D_{2}|}$, where $D_{1}$ is the proof of $s = t$ and $D_{2}$ is the proof of $t = u$ 
\item[] \translate{\OproveHyp} = \\${ProveHyp|x||y||D_{1}| (\lambda x:||\psi||.|D_{2}|)}$, where $D_{1}$ is the proof of $\varphi$ and $D_2$ is the proof of $\psi$. 
\end{itemize}

\textit{Sym}, \textit{Trans} and \textit{ProveHyp} have types as follows:

\begin{itemize}
\item[] $Sym: \Pi\alpha:type.\Pi x,y: term\, \alpha.\, proof(eq\,bool\,x\,y)\rightarrow proof(eq\,bool\,y\,x)$
\item[] $Trans: \Pi\alpha:type.\Pi x,y,z:term\,\alpha  \, proof(eq\,\alpha\,x\,y)\rightarrow proof(eq\,\alpha\,y\,z)\rightarrow proof(eq\,\alpha\,x\,z)$
\item[] $
ProveHyp: \Pi x,y:term\,bool.proof\,x\rightarrow (proof\,x\rightarrow\,proof\,y )\rightarrow proof\,y$
\end{itemize}

 \begin{figure}[]
\caption{A Proof Checking Page of ProofCloud}
\centering
\includegraphics[width=0.95\textwidth]{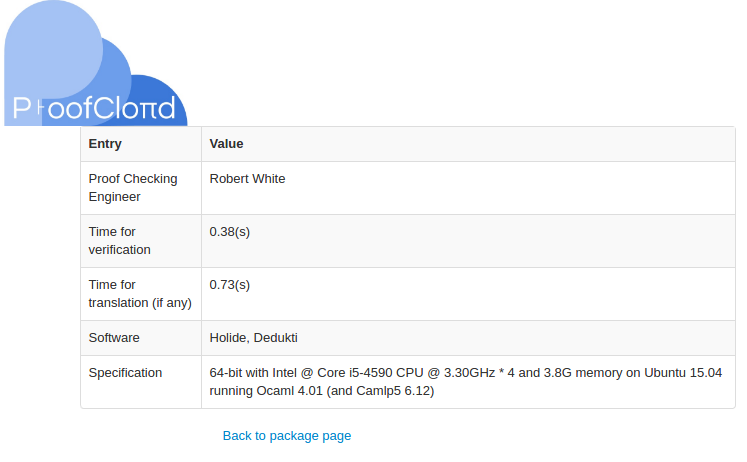}
\label{pkg}
\end{figure}

\section{Evaluation and Benchmarks}
\label{ch:eval}

ProofCloud includes also the proof checking results for all packages of OpenTheory. The standard library of OpenTheory grouped theorems into packages, including theorems of booleans, sets, lists, etc. The standard library (the base package, including 11 subpackages) was first verified in this project, followed by all the packages in the OpenTheory repository. For the first time, Holide and Dedukti successfully translated and  checked all packages in OpenTheory. Table \ref{tb:size1} illustrates the size of OpenTheory proof article files and the time taken for translation. Table \ref{tb:size2} represents the size of translated files and the time taken for proof checking by Dedukti. We name the translators as Holide 1 and Holide 2 respectively corresponding to OpenTheory version 5 and 6 respectively in Table \ref{tb:size1} and Table \ref{tb:size2}. Both article files and Dedukti files are compressed by $gzip$ to reduce the effect of syntax formatting and whitespace.  These benchmarks were generated on a 64-bit Intel Core i5-4590 CPU @3.30GHz $\times$4 PC with 3.8GB RAM. However, there is little difference in terms of the size of article files and the efficiency  between OpenTheory 5 and 6 for proof checking. The size of proof articles were reduced by around 7\% while the proof checking time was reduced by around 5\%. Figure \ref{pkg} gives an example of the proof checking page of the stream package.

\begin{table*}[]
\centering
\caption{Size of Article Files and Translation Time}
\label{tb:size1}
\begin{tabular}{|l|l|l|l|l|}
\cline{1-5}
\multirow{2}{*}{Package}                       & \multicolumn{2}{l|}{{ Holide 1}} & \multicolumn{2}{l|}{{ Holide 2}} \\ \cline{2-5} 
                                        & Size (KB)     & Time (s)       & Size (KB)     & Time (s)       \\ \hline
\multicolumn{1}{|l|}{base}              & 1,436             & 19.35          & 1,194              & 19.42          \\ \hline
\multicolumn{1}{|l|}{cl}                & 313              & 5.77           & 313             & 5.56           \\ \hline
\multicolumn{1}{|l|}{empty}             & 0                  & 0.20           & 0                  & 0.00           \\ \hline
\multicolumn{1}{|l|}{gfp}               & 136                & 1.42           & 112                & 1.35           \\ \hline
\multicolumn{1}{|l|}{lazy-list}         & 1,390              & 31.43          & 1,391              & 31.78          \\ \hline
\multicolumn{1}{|l|}{modular}           & 45                & 1.13           & 37                & 0.37           \\ \hline
\multicolumn{1}{|l|}{natural-bits}      & 162               & 1.43           & 132                & 1.39           \\ \hline
\multicolumn{1}{|l|}{natural-divides}   & 193              & 2.10           & 157              & 1.94           \\ \hline
\multicolumn{1}{|l|}{natural-fibonacci} & 130                & 1.31           & 108                & 1.24           \\ \hline
\multicolumn{1}{|l|}{natural-prime}     & 140                & 1.46           & 116                & 1.34           \\ \hline
\multicolumn{1}{|l|}{parser}            & 240              & 3.22           & 204              & 3.15           \\ \hline
\multicolumn{1}{|l|}{probability}       & 26                & 0.30           & 23                & 0.23           \\ \hline
\multicolumn{1}{|l|}{stream}            & 75                & 0.75           &  63               & 0.73           \\ \hline
\multicolumn{1}{|l|}{word10}            & 86                & 0.76           & 71                & 0.62           \\ \hline
\multicolumn{1}{|l|}{word12}            & 88                & 0.79           & 72                & 0.75           \\ \hline
\multicolumn{1}{|l|}{word16}            & 131                & 1.60           & 107                & 0.77           \\ \hline
\multicolumn{1}{|l|}{word5}             & 77                & 0.70           & 64                & 1.56           \\ \hline
\multicolumn{1}{|l|}{{\bf Total}}       & {\bf 4,668}       & {\bf 73.73}    & {\bf 4,377}       & {\bf 72.21}    \\ \hline
\end{tabular}
\end{table*}

\begin{table*}[]
\centering
\caption{Size of Dedukti Files and Proof Checking Time}
\label{tb:size2}
\begin{tabular}{|l|l|l|l|l|}
\cline{1-5}
\multirow{2}{*}{Package}                       & \multicolumn{2}{l|}{{Dedukti (Holide 1)}} & \multicolumn{2}{l|}{{Dedukti (Holide 2)}} \\ \cline{2-5} 
                                        & Size (KB)     & Time (s)       & Size (KB)     & Time (s)       \\ \hline
\multicolumn{1}{|l|}{base}              & 4,681             & 10.63          & 4,440             & 9.74           \\ \hline
\multicolumn{1}{|l|}{cl}                & 1,219             & 2.42           & 1,219             & 2.46           \\ \hline
\multicolumn{1}{|l|}{empty}             & 0                  & 0.00           & 0                  & 0.00           \\ \hline
\multicolumn{1}{|l|}{gfp}               & 400              & 0.73           & 375              & 0.65           \\ \hline
\multicolumn{1}{|l|}{lazy-list}         & 5,718             & 13.31          & 5,717             & 13.11          \\ \hline
\multicolumn{1}{|l|}{modular}           & 120              & 0.19           & 111              & 0.17           \\ \hline
\multicolumn{1}{|l|}{natural-bits}      & 452              & 0.74           & 419              & 0.68           \\ \hline
\multicolumn{1}{|l|}{natural-divides}   & 599              & 1.11           & 566              & 0.99           \\ \hline
\multicolumn{1}{|l|}{natural-fibonacci} & 378              & 0.67           & 354              & 0.60           \\ \hline
\multicolumn{1}{|l|}{natural-prime}     & 408              & 0.72           & 388              & 0.65           \\ \hline
\multicolumn{1}{|l|}{parser}            & 802              & 1.87           & 776              & 1.69           \\ \hline
\multicolumn{1}{|l|}{probability}       & 72                & 0.12           & 69                & 0.11           \\ \hline
\multicolumn{1}{|l|}{stream}            & 221              & 0.41           & 211              & 0.38           \\ \hline
\multicolumn{1}{|l|}{word10}            & 234              & 0.38           & 216              & 0.29           \\ \hline
\multicolumn{1}{|l|}{word12}            & 239              & 0.40           & 220              & 0.35           \\ \hline
\multicolumn{1}{|l|}{word16}            & 396              & 0.80           & 364              & 0.36           \\ \hline
\multicolumn{1}{|l|}{word5}             & 207              & 0.33           & 192              & 0.72           \\ \hline
\multicolumn{1}{|l|}{{\bf Total}}       & {\bf 16,146}      & {\bf 34.83}    & {\bf 15,637}      & {\bf 32.95}    \\ \hline
\end{tabular}
\end{table*}

\section{Conclusion}

We have introduced ProofCloud and explained the implementation. ProofCloud is a customised search engine providing retrieval service for mathematicians and computer scientists using OpenTheory and reasoners in the HOL family. While OpenTheory packs proofs up, ProofCloud unpacks them and displays the proofs on webpages. It also help promoting OpenTheory and improve the usability of it. ProofCloud also presents proof checking results. The previous version of Holide can only translate the standard library of OpenTheory \cite{holide}. For the first time, all OpenTheory packages were translated by Holide and passed proof checking. This also provides evidence that OpenTheory is a reliable platform for higher order proofs and has validated the updates of Holide. However, as we have seen, there is little impact on the size of article files and efficiency of proof checking. Future work includes adding the remaining packages of OpenTheory to ProofCloud. Most recent updates of OpenTheory include also the names of proofs in packages. This would benefit the usability of  ProofCloud.

\section{Acknowledgement}

The author was supported by MPRI-INRIA scholarship for this internship. During this project, the author received kind help from Prof. Gilles Dowek, Dr. Ali Assaf, Dr. Joe Hurd, and Mr. Fr\'ed\'eric Gilbert on the understanding and implementation of Holide, OpenTheory, and ProofCloud.

% \nocite{*}
% \bibliographystyle{eptcs}

\appendix

\section{The Specification of ProofCloud}
\label{spec}
The following attributes are included in the webpages of each package:

% Further contribution

\begin{multicols}{2}
\renewcommand{\baselinestretch}{0.3}
\begin{itemize}
\item Package name
\item Author of package
\item Subpackages
\item Date retrieved
\item Total number of proofs
\item Number of constructive proofs
\item Number of classical proofs
\item Percentage of constructive proofs
\item Size of constructive proofs on average
\item Size of classical proofs on average
\item List of proofs (names and conclusions)
\item Comments
\end{itemize}
\end{multicols}

For each package, there is also a page for verification (proof checking) information with the following entries:

\begin{multicols}{2}
\renewcommand{\baselinestretch}{0.3}
\begin{itemize}
\item Software engineer for verification
\item Software for verification
\item Translation time
\item Verification time
\item PC Specification
\item Comments
\end{itemize}
\end{multicols}

Each proof has its own page for structural, statistical and proof checking results with the attributs as follows:

\begin{multicols}{2}
\begin{itemize}
\item Theorem name
\item Theorem conclusion
\item Packagename
\item Constructive proof (or not)
\item Axioms
\item Constructive lemmas (if any)
\item Classical lemmas (if any)
\item Package
\item Comments
\end{itemize}
\end{multicols}

\bibliographystyle{unsrt}
\bibliography{main}

\end{document}